\documentclass[12pt]{article}

\usepackage{cite}
\usepackage{subfigure}
\usepackage{multirow}
\usepackage{helvet}
\usepackage{amsmath}
\usepackage{amssymb}
\usepackage{setspace}
\usepackage{setspace}
\usepackage[dvips]{graphicx}
\usepackage{epsfig}
\usepackage{empheq}

\begin{document}

\titlepage                                                    
 \begin{flushright}                                                    
 IPPP/14/80  \\
 DCPT/14/160 \\                                                                                                       
  \end{flushright} 
 \vspace*{0.5cm}
\begin{center}                                                    
{\Large \bf Exclusive production of double $J/\psi$ mesons in\\[0.2 cm] hadronic collisions}\\

\vspace*{1cm}
                                                   
L.A. Harland--Lang$^{1}$, V.A. Khoze$^{1,2}$, M.G. Ryskin$^{2}$ \\                                                 
                                                   
\vspace*{0.5cm}                                                    
${}^1$ Institute for Particle Physics Phenomenology, University of Durham, Durham, DH1 3LE          \\
${}^2$ Petersburg Nuclear Physics Institute, NRC Kurchatov Institute, Gatchina, \linebreak[4]St. Petersburg, 188300, Russia                                              
                                                    
\vspace*{1cm}                         

\begin{abstract}

\noindent We present the first calculation of exclusive double $J/\psi$ production in hadronic collisions. We analyse in detail the form of the Born--level $gg \to J/\psi J/\psi$ amplitudes within the non--relativistic quarkonium approximation and discuss the implications of this for the central exclusive production channel, within the `Durham' perturbative model. In addition we show that this direct single parton scattering contribution is expected to be strongly dominant in the exclusive case. We present predictions for the LHC and show that the expected cross sections are in fair agreement with the LHCb Run--I  measurement of exclusive double $J/\psi$ production, with the measured invariant mass distribution described well by the theory. Motivated by this encouraging result we present predictions for observables that may be measured in LHC Run--II, and estimate the size of the expected cross sections in the $\psi(2S)$ and $\chi_{c}$ cases.

\end{abstract}
                                   
\end{center}  

\section{Introduction}

Central exclusive production (CEP) is the process
\begin{equation}\label{exc}
pp \to p+X+p\;,
\end{equation}
where `$+$' signs are used to denote the presence of large rapidity gaps, separating the system $X$ from intact outgoing protons. Such reactions represent an experimentally very clean signal and provide a promising way to investigate both QCD dynamics and new physics in hadronic collisions, see~\cite{Harland-Lang:2014lxa,Harland-Lang:2014dta,Albrow:2010yb} for recent reviews. From a theoretical point of view, these processes are interesting because they involve both soft, diffractive physics, as indicated by the exchange of vacuum quantum numbers between the colliding hadrons and the corresponding presence of large rapidity gaps, and hard physics via the production process of the central particle. Moreover, it is found that a dynamical selection rule operates, where $J_z^{PC}=0^{++}$ quantum number states (here $J_z$ is the projection of the object angular momentum on the beam axis) are expected to be dominantly produced; this simple fact leads to many interesting and non--trivial implications which are not seen in the inclusive case.

The inclusive production of charmonia pairs in hadronic collisions is a highly topical subject, see for example~\cite{Qiao:2009kg,Kom:2011bd,Novoselov:2011ff,Baranov:2011ch,Lansberg:2013qka,Lansberg:2014swa} for recent theoretical work. Such a process is of interest both as a test of the tools of heavy quarkonium theory, but also because the probability that two $J/\psi$ mesons can be produced in independent scatters may be quite large at the LHC, due to the large cross section for inclusive single $J/\psi$ production and the large flux of incoming partons. Inclusive double $J/\psi$ production has been observed by LHCb~\cite{Aaij:2011yc} and CMS~\cite{CMS:2013pph} at the LHC, and there are intriguing hints in the data of a double parton scattering (DPS) contribution: in the LHCb case the measured $J/\psi J/\psi$ invariant mass distribution may be somewhat broader than expected from the single parton scattering (SPS) $gg \to J/\psi J/\psi$ process, while CMS observe an excess at higher rapidity difference $\Delta y$ between the $J/\psi$ mesons, where DPS contributions are expected to be important. This was confirmed by a more recent D0 analysis~\cite{Abazov:2014qba} at the Tevatron,  where a clear excess at large $\Delta y$ was seen, allowing the DPS contribution to be separated and measured (see also~\cite{Lansberg:2014swa}). The possibility of a contribution to this signal from the decay of exotic particles such as tetra--charm--quarks has also been discussed in~\cite{Berezhnoy:2011xy}, while the conventional $\eta_b$ is also expected to decay to two $J/\psi$ mesons.

More recently \emph{exclusive} double $J/\psi$ and $J/\psi \psi(2S)$ production has been observed by LHCb~\cite{Aaij:2014rms}, by selecting events with no additional activity in the LHCb acceptance, which is sensitive to charged particles in the pseudorapidity ranges $-3.5<\eta<-1.5$ and $1.5<\eta<5.0$, and applying a data--driven correction to subtract the contribution of events where additional particles due to e.g. proton dissociation, lie outside this acceptance. This process is of great theoretical interest as a test of the underlying perturbative formalism and its various non--trivial ingredients, and moreover as we will demonstrate in this paper, the probability for the charmonia to be produced exclusively in independent scatters is very small; thus the direct $gg \to J/\psi J/\psi$ production channel is dominantly probed. In addition, such exclusive reactions are purely sensitive to the colour--singlet component of the meson wave function: they do not receive colour--octet contributions, for which additional radiation will be present. These reactions are also in principle sensitive to additional particles which might be produced in decay chains that involve exotic particles.

In this paper we present the first calculation of exclusive double $J/\psi$ and $\psi(2S)$ production (i.e. with $X=J/\psi J/\psi$, $J/\psi \psi(2S)$ or $\psi(2S) \psi(2S)$ in (\ref{exc})) in hadronic collisions. We will see that the predicted cross sections are in reasonable agreement with the LHCb measurements, within theoretical and experimental uncertainties, while the shape of the measured $J/\psi J/\psi$ invariant mass distribution is well described, with no hint of the discrepancy discussed above, which may be present in the inclusive case. While this lends some support to the perturbative CEP framework, we will also comment on the possibilities for better testing the theory by for example considering ratios of various observables. In addition, we will present a detailed analysis of the contributing $gg \to J/\psi J/\psi$ helicity amplitudes, which may be of more general interest, in particular in the inclusive channel.

The outline of this paper is as follows. In Section~\ref{durt} we briefly describe the `Durham' perturbative model for the CEP process. In Section~\ref{jpsisub} we present results for the $gg \to J/\psi J/\psi$ helicity amplitudes, considering their explicit analytic form (within the non--relativistic approximation and to leading order in $\alpha_s$) in the threshold and high subprocess energy limits, and the implications for the exclusive production channel. In Section~\ref{secsym} we consider the double photoproduction and `symmetric' production mechanism, where two separate but kinematically correlated $gg \to c\overline{c}$ scatters occur, and show that these contributions are expected to be very small. In Section~\ref{res} we present numerical predictions for exclusive double $J/\psi$ production at the LHC, as well as giving estimates for $\psi(2S)J/\psi$ and double $\psi(2S)$ production, and discussing the possibilities for double $\chi_c$ and $\eta_c$ production. Finally in Section~\ref{conc} we conclude.

\section{Central exclusive production}\label{durt}

\begin{figure}
\begin{center}
\includegraphics[scale=1.2]{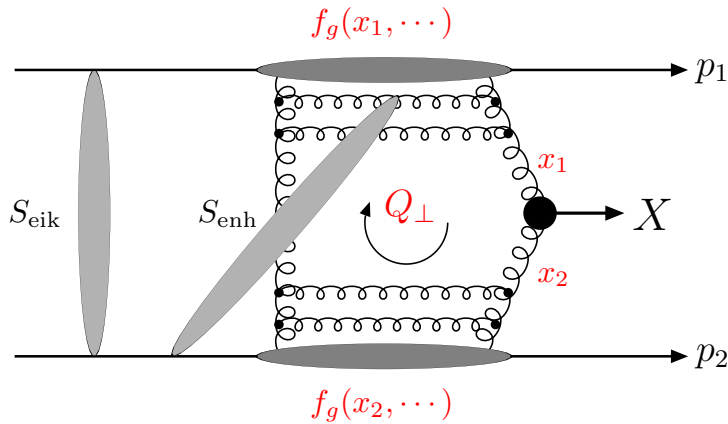}
\caption{The perturbative mechanism for the exclusive process $pp \to p\,+\, X \, +\, p$, with the eikonal and enhanced survival factors 
shown symbolically.}
\label{fig:pCp}
\end{center}
\end{figure} 

The `Durham' model is a pQCD--based approach to modelling the CEP process (\ref{exc}), when the object mass $M_X$ is sufficiently high. It represents a novel application of pQCD, as well as requiring an account of soft diffractive physics: see~\cite{Harland-Lang:2014lxa} for further discussion. The formalism used to calculate the perturbative CEP cross section is explained in detail elsewhere~\cite{Khoze97,Khoze00,Kaidalov03,HarlandLang:2010ep,Khoze:2001xm,Albrow:2010yb,Harland-Lang:2013xba} and we will only present a very brief summary here. The perturbative CEP amplitude, corresponding to the diagram shown in Fig.~\ref{fig:pCp}, can be written as
\begin{equation}\label{bt}
A=\pi^2 \int \frac{d^2 {\bf Q}_\perp\, \overline{\mathcal{M}}}{{\bf Q}_\perp^2 ({\bf Q}_\perp-{\bf p}_{1_\perp})^2({\bf Q}_\perp+{\bf p}_{2_\perp})^2}\,f_g(x_1,x_1', Q_1^2,\mu_F^2; t_1)f_g(x_2,x_2',Q_2^2,\mu_F^2;t_2)\; ,
\end{equation}
where  $Q_\perp$ is the transverse momentum in the gluon loop, with the scale $Q_i^2 = Q_\perp^2$ in the forward proton limit (see e.g.~\cite{HarlandLang:2010ep} for a prescription away from this limit) and $\overline{\mathcal{M}}$ is the colour--averaged, normalised sub--amplitude for the $gg \to X$ process
\begin{equation}\label{Vnorm}
\overline{\mathcal{M}}\equiv \frac{2}{M_X^2}\frac{1}{N_C^2-1}\sum_{a,b}\delta^{ab}q_{1_\perp}^\mu q_{2_\perp}^\nu V_{\mu\nu}^{ab} \; .
\end{equation}
Here $a$ and $b$ are colour indices, $M_X$ is the central object mass, $V_{\mu\nu}^{ab}$ is the $gg \to X$ vertex, $q_{i_\perp}$ are the transverse momenta of the incoming gluons, and $t_i$ is the squared momentum transfer to the outgoing protons. The $f_g$'s in (\ref{bt}) are the skewed unintegrated gluon densities of the proton. These correspond to the distribution of gluons in transverse momentum $Q_\perp$, which are evolved in energy up to the hard scale $\mu_F$, such that they are accompanied by no additional radiation, as is essential for exclusive production. It can be shown that, in the kinematic regime relevant to CEP, the $f_g$'s can be written as
\begin{equation}\label{fgskew}
f_g(x,x',Q_\perp^2,\mu_F^2) =\; \frac{\partial}{\partial \ln(Q_\perp^2)} \left[ H_g\left(\frac{x}{2},\frac{x}{2};Q_\perp^2\right) \sqrt{T_g(Q_\perp,\mu_F^2)} \right]\;,
\end{equation}
where $H_g$ is the generalised gluon PDF~\cite{Belitsky:2005qn}, which for CEP kinematics can be related to the conventional PDFs~\cite{Shuvaev:1999ce,Harland-Lang:2013xba}. The $T_g$ in (\ref{fgskew}) is a Sudakov factor, which corresponds to the probability of no extra parton emission from each fusing gluon. 

In addition to the perturbative amplitude (\ref{bt}), we must also include the probability that extra particles are not produced in additional soft proton--proton interactions (or `rescatterings'), largely independent of the hard process, i.e. as a result of underlying event activity. This probability is encoded in the so--called `eikonal survival factor', $S^2_{\rm eik}$~\cite{Bjorken:1992er,Khoze:2002nf,Martin:2009ku,Gotsman:2014pwa}. While the survival factor is a soft quantity which cannot be calculated using pQCD, it may be extracted from hadronic data~\cite{Khoze:2013dha,Khoze:2013jsa}. Although there is some uncertainty in the precise level of suppression (in particular in its dependence on the c.m.s. energy $\sqrt{s}$), it is found to be a sizeable effect, reducing the CEP cross section by about two orders of magnitude. An additional, `enhanced' survival factor~\cite{Ryskin:2009tk,Martin:2009ku,Ryskin:2011qe,Gotsman:2014pwa}, which corresponds to a suppression caused by the rescatterings of the protons with the intermediate partons, should also in general be included, although it is expected to be a much less significant effect.  In this paper we will consider different versions of the `two--channel' eikonal model of~\cite{Khoze:2013dha}: the parameters are tuned to give a satisfactory reproduction of the elastic and diffractive dissociation data observed at the LHC and the lower energy colliders. For simplicity, we will omit any enhanced absorptive effects from our calculation.

To calculate the cross section for exclusive double $J/\psi$ production, we can decompose (\ref{Vnorm}) in terms of on--shell helicity amplitudes, neglecting small off--shell corrections of order $\sim q_\perp^2/M_X^2$. Omitting colour indices for simplicity, this gives
\begin{align}
q_{1_\perp}^i q_{2_\perp}^j V_{ij} =\begin{cases} &-\frac{1}{2} ({\bf q}_{1_\perp}\cdot {\bf q}_{2_\perp})(T_{++}+T_{--})\;\;(J^P_z=0^+)\\ 
&-\frac{i}{2} |({\bf q}_{1_\perp}\times {\bf q}_{2_\perp})|(T_{++}-T_{--})\;\;(J^P_z=0^-)\\ 
&+\frac{1}{2}((q_{1_\perp}^x q_{2_\perp}^x-q_{1_\perp}^y q_{2_\perp}^y)+i(q_{1_\perp}^x q_{2_\perp}^y+q_{1_\perp}^y q_{2_\perp}^x))T_{-+}\;\;(J^P_z=+2^+)\\ 
&+\frac{1}{2}((q_{1_\perp}^x q_{2_\perp}^x-q_{1_\perp}^y q_{2_\perp}^y)-i(q_{1_\perp}^x q_{2_\perp}^y+q_{1_\perp}^y q_{2_\perp}^x))T_{+-}\;\;(J^P_z=-2^+)
\end{cases}\label{Agen}
\end{align}
where the $J_z^P$ indicate the parity and spin projection on the $gg$ axis, and $T_{\lambda_1\lambda_2}$ are the corresponding $g(\lambda_1)g(\lambda_2)\to X$ helicity amplitudes, see~\cite{HarlandLang:2010ep,Harland-Lang:2014lxa} for more details. In the forward proton limit the only non-vanishing term after the $Q_\perp$ integration (\ref{bt}) is the first one: this is the origin of the selection rule~\cite{Kaidalov:2003fw,Khoze:2000mw,Khoze:2000jm} which operates in this exclusive process, and strongly favours $J^P_z=0^+$ quantum numbers for the centrally produced state. More generally, away from the exact forward limit the non-$J^P_z=0^+$ terms in (\ref{Agen}) do not give completely vanishing contributions to the $Q_\perp$ integral and we find that 
\begin{equation}\label{simjz2}
\frac{|A(|J_z|=2)|^2}{|A(J_z=0)|^2} \sim \frac{\langle p_\perp^2 \rangle^2}{\langle Q_\perp^2\rangle^2}\;,
\end{equation}
 which is typically of order $\sim1/50-1/100$, depending on such factors as the central object mass, cms energy $\sqrt{s}$ and choice of PDF set~\cite{HarlandLang:2010ep,Harland-Lang:2014lxa}.

\section{The $gg \to J/\psi J/\psi$ subprocess}\label{jpsisub}

\begin{figure}
\begin{center}
\includegraphics[scale=0.6]{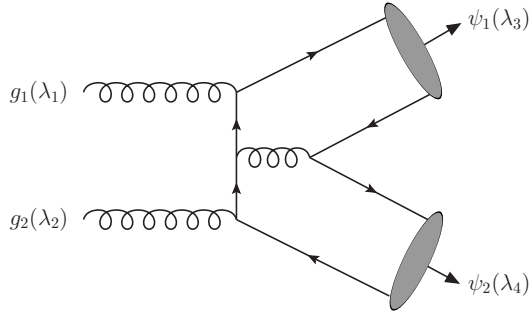}
\caption{Typical Feynman diagram contributing to $gg \to J/\psi J/\psi$ process.}\label{feyn}
\end{center}
\end{figure}

To calculate the cross section for exclusive double $J/\psi$ production, we need expressions for the $gg \to J/\psi J/\psi$ helicity amplitudes, which can then be used in (\ref{Agen}) to arrive at a result for the CEP cross section. A typical diagram for this process is shown in Fig.~\ref{feyn}; altogether there are 31 separate Feynman diagrams to consider. Explicit expressions for the $gg\to VV$ helicity amplitudes, for the production of massless vector mesons (i.e. relevant for the case that $M_X \gg M_\psi$) are given in~\cite{HarlandLang:2011qd}, and this calculation could readily be extended to the case of a non--zero quark mass, in the non--relativistic limit. However, an explicit expression for the tensor $M^{\mu\nu\rho\sigma}_{ab}$ that is contracted with the incoming gluon and outgoing $J/\psi$ polarization vectors is given in~\cite{Qiao:2009kg}, for the case of the non--relativistic colour--singlet model and to LO in $\alpha_s$. This is precisely the object needed in the case of CEP, and we will make use of this result throughout this paper.

\subsection{High energy limit}

Although the tensor $M^{\mu\nu\rho\sigma}_{ab}$ in~\cite{Qiao:2009kg} is in general quite a complicated object, in the high--energy ($M_X \gg M_V$) limit effect from the non--zero charm mass will become sub--leading, and this should reduce to the relatively simple expressions given in~\cite{HarlandLang:2011qd} for the production of massless vector meson pairs\footnote{As discussed in~\cite{HarlandLang:2011qd}, the massless $gg\to VV$ amplitudes receive no contribution from the `ladder--type' diagrams, where the $q\overline{q}$ pairs forming the meson come from the same quark line: these cannot contribute in the case of $J/\psi$ production due to $C$--parity conservation.}. In this case the calculation was performed in the `hard exclusive' formalism of~\cite{Brodsky:1981kj,Chernyak:1981zz}, for which the quark, anti--quark caries some general momentum fraction $x$, $(1-x)$ of the parent meson, respectively: as we are considering the non--relativistic approximation (taken in e.g.~\cite{Qiao:2009kg,Kom:2011bd,Novoselov:2011ff,Baranov:2011ch,Lansberg:2013qka,Lansberg:2014swa} ), we simply assume $x=1/2$. In addition, to account for the different normalizations of the spin 
projections, we 
must make the replacement
\begin{equation}
f_M \to \sqrt{\frac{3}{\pi M_\psi}}R_0(0)\;,
\end{equation}
where $f_M$ is the meson decay constant, defined in~\cite{HarlandLang:2011qd}, and $R_0(0)$ is the $J/\psi$ wave function at the origin. Making these replacements we therefore expect the helicity amplitudes $T_{\lambda_1\lambda_2,\lambda_3\lambda_4}$ for transverse $J/\psi$ polarizations, in the $M_X^2/M_\psi^2 \to \infty$ limit, to be given by\footnote{We note that all of the explicit expressions presented in this paper and in~\cite{HarlandLang:2011qd,Harland-Lang:2013bya} correspond to taking the azimuthal angle $\phi=0$ for the outgoing particle momenta. Taking a non--zero value can introduce an overall $\phi$--dependent phase for some helicity configurations, which while having no effect in spin--summed the normal inclusive case, must be included in (\ref{Agen}) to give the correct result for the exclusive cross section.}
\begin{align}\label{Atran1}
T_{\lambda_1\lambda_2,++}&=T_{\lambda_1\lambda_2,--}=0\;,\\ \label{Atran2}
T_{++,+-}&=T_{++,-+}=T_{--,+-}=T_{--,-+}=0\;,\\ \label{Atran3}
T_{+-,+-}&=T_{-+,-+}=-\delta^{ab}\,\frac{8 |R_0(0)|^2}{\pi M_\psi}\, \frac{16\pi^2\alpha_s^2}{\hat{s}}\cos\theta(1+\cos\theta)\;,\\ \label{Atran4}
T_{-+,+-}&=T_{+-,-+}=\delta^{ab}\,\frac{8 |R_0(0)|^2}{\pi M_\psi}\, \frac{16\pi^2\alpha_s^2}{\hat{s}}\cos\theta(1-\cos\theta)\;,
\end{align}
where $\sqrt{\hat{s}}=M_X$ is the invariant mass of the $J/\psi$ pair, $\lambda_{1,2}$ are the helicities of the incoming gluons and $\lambda_{3,4}$ are the helicities of the outgoing $J/\psi$ mesons, while $a,b$ are the gluon colour indices\footnote{We can see from these expressions that the non--zero $gg \to J/\psi J/\psi$ amplitudes scale as $\sim 1/\hat{s}$ in the high energy limit, with similar results holding for longitudinal polarizations, and so we might naively expect a $\sim 1/\hat{s}^3$ scaling in the subprocess cross section. However, in fact some of the amplitudes contain terms which while strictly vanishing in the $M_\psi \to 0$ limit, have near singularities as $|\cos\theta|\to 1$ that are regulated by the non--zero $J/\psi$ mass. This actually results in a $\sim 1/\hat{s}^2$ scaling in the high energy limit, as was found in~\cite{Berezhnoy:2011xy}, when the cross section is integrated over \emph{all} $\cos\theta$. However, as soon as any cut is placed to remove the $|\cos\theta|\to 1$ forward phase space region, the expected $\sim 1/\hat{s}^3$ scaling is restored, and the expressions (\ref{Atran1})--(\ref{Along3}) can be used: as such a cut will always be placed experimentally, it is this scaling that is phenomenologically relevant, in both the exclusive and inclusive cases.}. Here (\ref{Atran1}) simply results from helicity conservation along the massless quark lines, while (\ref{Atran2}) follows from the well--known fact that a tree--level amplitude for the scattering of massless particles must have at least 2 particles of both $+$ and $-$ helicity (with all particle momenta defined as incoming) to be non--zero~\cite{Mangano:1990by}. For the case that exactly two particles have the same helicity the amplitudes are `maximally helicity violating' (MHV), and in such a situation very simple expressions for these can be written down~\cite{Parke:1986gb,Berends:1987me,Mangano:1990by}. It is precisely this simplicity that explains why the expressions (\ref{Atran3}) and (\ref{Atran4}) are so concise; recalling that these expressions result from the contribution of $31$ separate Feynman diagrams~\cite{HarlandLang:2011qd}, this would not in general be expected.

Considering now the case of longitudinal polarizations, we expect
\begin{align}\label{Along1}
T_{\lambda_1\lambda_2,0\pm}&=T_{\lambda_1\lambda_2,\pm 0}=0\;,\\ \label{Along2}
T_{++,00}&=T_{--,00}=0\;,\\ \label{Along3}
T_{+-,00}&=T_{-+,00}=-\delta^{ab}\,\frac{8 |R_0(0)|^2}{\pi M_\psi}\, \frac{16\pi^2\alpha_s^2}{\hat{s}}\left(\cos^2\theta-\frac{C_F}{N_c}\right)\;.
\end{align}
Here (\ref{Along2}) again simply follows from helicity conservation along the massless quark line. The expression (\ref{Along3}) has the remarkable feature, first observed in~\cite{HarlandLang:2011qd} for the case of $\pi\pi$ production, that it vanishes for a particular value of $\cos^2\theta$ (corresponding to $\theta\approx\pm 48^\circ$). This vanishing of a Born amplitude for the radiation of massless gauge bosons, for a certain configuration of the final state particles, is a known effect, usually labeled a `radiation zero', see for instance~\cite{Heyssler:1997ng,Brown:1982xx}.  While this effect, which is present in all theories with massless gauge bosons, is expected to occur in QCD, it is usually neutralised along with colour by the averaging of hadronisation. Double $J/\psi$ production therefore presents an interesting possibility to observe such a QCD radiation zero, although an experimentally challenging one, in particular as it is not easy to select dominantly longitudinally polarized $J/\psi$ mesons. We note that this possibility is not limited to the purely exclusive mode: as the amplitudes (\ref{Along3}) are the only non--zero ones, such a radiation zero should also be present at high enough $\hat{s}$ inclusively, for longitudinal $J/\psi$ polarizations, although in this case there may be colour octet (as well as higher--order) contributions to consider, which would not be expected to exhibit such zeros.

We have confirmed by taking the result in~\cite{Qiao:2009kg} and making the relevant replacements discussed above that this does indeed reduce to these simple results in the $\hat{s}/M_\psi^2 \to \infty$ limit. Thus, in the high--energy limit only a small number of the possible helicity amplitudes contribute, with remarkably simple forms. Moreover, as we can see that the amplitudes with $J_z=0$ incoming gluons all vanish, we will expect in this asymptotic region a strong suppression in the CEP of $J/\psi J/\psi$ mesons, due to the selection rule discussed above, which disfavours non--$J_z^P=0^+$ quantum numbers of the centrally produced state, see (\ref{simjz2}). Away from this exact high--energy limit mass corrections will enter and the resulting amplitudes will not be so simple, with the $J_z=0$ amplitudes in general being non-vanishing, although we can nevertheless expect these corrections to be small sufficiently far away from threshold. However in the lower mass region we should not expect to trust these simple expressions: in the following section, we therefore consider the opposite, threshold, limit and show again that quite simple expressions can be written down for the corresponding amplitudes.

\subsection{Threshold limit}\label{sec:thres}

\begin{figure}
\begin{center}
\includegraphics[scale=0.65]{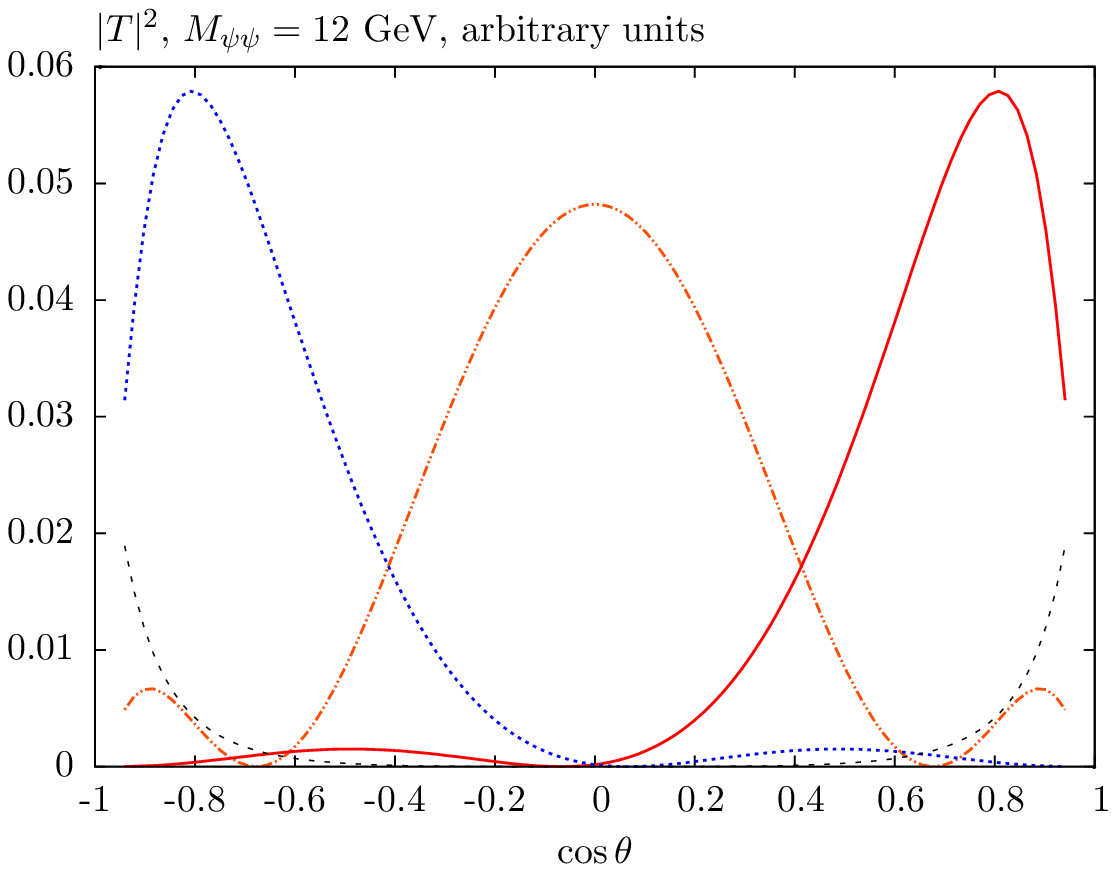}\qquad
\includegraphics[scale=0.65]{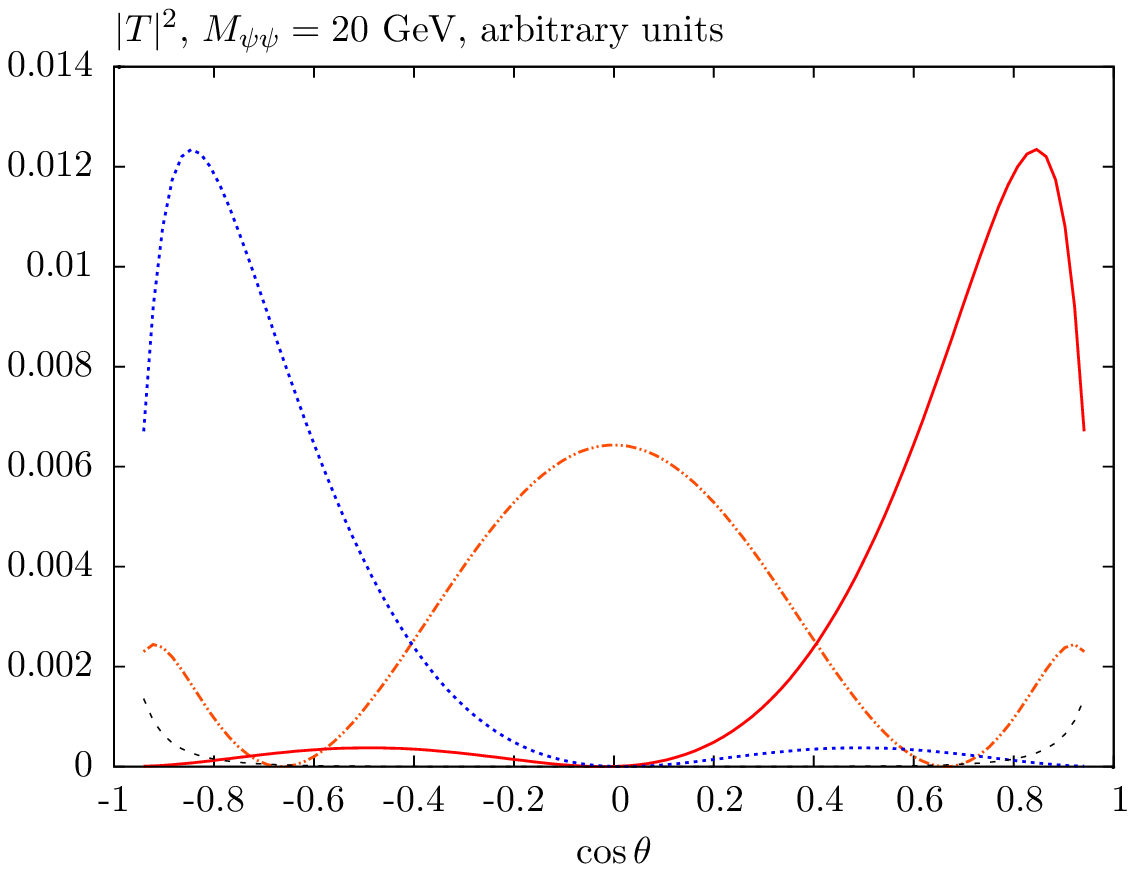}
\caption{Angular distribution of the squared helicity amplitudes for $g(\lambda_1)g(\lambda_2) \to J/\psi(\lambda_3) J/\psi(\lambda_4)$ process, in arbitrary units, for $M_{\psi \psi}$=12, 20 GeV. The solid, dotted, dashed and dot--dashed curves correspond to $(\lambda_1\lambda_2\lambda_3\lambda_4)=+-+-$, $+--+$, $++00$ and $+-00$, respectively. The $+-+-$ and $+--+$ curves have been divided by 4 for display purposes.}\label{ang}
\end{center}
\end{figure}

In the threshold ($s \approx 4 M_\psi^2$) limit, we find that the $gg \to J/\psi J/\psi$ amplitudes also take very simple forms: these are given for general polarizations in Appendix~\ref{app:thres}. Taking some representative amplitudes, we have
\begin{align}\label{thresrep1}
T_{+++-}&=T_{+++0}=0\;,\\\label{thresrep2}
T_{++++}&=T_{++00}= \frac{8\pi\alpha_s^2|R_0(0)|^2}{9 M_\psi^3}\;,\\\label{thresrep3}
T_{+-++}&=T_{+---}=- \frac{8\pi\alpha_s^2|R_0(0)|^2}{9 M_\psi^3}\,7(1-\cos^2\theta)\;,
\end{align}
with similar results holding for other polarizations. We therefore find that for example
\begin{equation}\label{supthres}
\frac{|T_{++++}|^2}{|T_{+-++}|^2}=\frac{1}{49}\;,
\end{equation}
for $90^\circ$ scattering; for all other non--zero amplitudes with $J_z=0$ incoming gluons a similar or stronger suppression is seen. We therefore find a strong numerical suppression in the amplitudes for which the gluons are in a $J_z=0$ configuration, and therefore in the $J/\psi J/\psi$ CEP cross section near threshold; in this lower mass region, both the numerically suppressed $J_z=0$ and dynamically suppressed $|J_z|=2$ configurations will contribute. Such a suppression in the $J_z=0$ amplitudes appears to be due entirely to the specific internal structure of the $J/\psi$ state, and is not seen in for example the equivalent $\gamma \gamma \to W^+W^-$ amplitudes~\cite{Belanger:1992qi}, in the threshold region.

We note that the vanishing of the Born amplitude $T_{+++-}$ (\ref{thresrep1}) is in fact found to hold in general\footnote{This applies equally to the related amplitudes with $\lambda_i \to -\lambda_i$ and/or $\lambda_3 \leftrightarrow \lambda_4$.}, that is for arbitrary $\hat{s}$. While for massless particles this amplitude must vanish at LO from MHV considerations~\cite{Parke:1986gb,Berends:1987me,Mangano:1990by}, it is not superficially obvious that this should be the case here, where the final--state particles have a mass. However, remarkably, this result is also found to hold in the case of the $\gamma \gamma \to W^+W^-$~\cite{Belanger:1992qi}, $\gamma\gamma \to q\overline{q}$ and $gg \to q\overline{q}$ Born amplitudes (with arbitrary quark masses, see e.g.~\cite{Khoze:2006um}\footnote{Although~\cite{Khoze:2006um} only considers the case that the gluons are in a colour--singlet state, it can readily be shown that this $gg \to q\overline{q}$ Born helicity amplitude vanishes for arbitrary colour.}). In for example~\cite{Borden:1994fv,Khoze:2006um} it was discussed how this vanishing could be shown to follow in the $\gamma\gamma \to q\overline{q}$ case from CPT invariance, photon Bose statistics and unitarity (the latter condition results in the invariance of the amplitude under time reversal, which only holds for the Born amplitude): we expect a similar argument to hold here.

Finally, in Fig.~\ref{ang} we show the angular distribution of the squared matrix elements for representative values of $M_X$ ($=12,20$ GeV). The dip structure in $+-00$ helicity amplitude, due to the radiation zero discussed above, is evident, although for the experimentally most realistic values of $M_X$ it is not too pronounced (we note however, as discussed above, that this dip survives in the case of inclusive production): the position of the dip approaches the value expected from (\ref{Along3}) as $M_X$ increases. It is also interesting to see that the $+-+-$ and $+--+$ amplitudes also exhibit zeros, at values of $\cos\theta$ which are slightly displaced from the $\cos\theta=0$ that is expected in the high--energy limit, see (\ref{Atran3},\ref{Atran4}). The dominance of the amplitudes with $|J_z|=2$ incoming gluons (divided by 4 in the figure for display purposes), expected from the discussion in the preceding sections, is also clear.

\section{`Symmetric' production process}\label{secsym}

As discussed in the introduction, in the case of \emph{inclusive} double $J/\psi$ pair production it is found that the contribution from DPS, where two $J/\psi$ meson are produced in independent scatters, may be quite large at the LHC, and may contribute to the observed LHCb~\cite{Aaij:2011yc} and CMS~\cite{CMS:2013pph} data, while a clear contribution is seen in the D0 data~\cite{Abazov:2014qba}. However, in the exclusive case single $J/\psi$ production cannot proceed via the type of digram shown in Fig.~\ref{fig:pCp}, due to the odd--$C$ parity of the meson. Rather, the lowest--order process proceeds via photoproduction, where the $J/\psi$ couples to a two--gluon $t$--channel exchange with one proton, and a photon exchange with the other. To give an estimate of the cross section for this process we can assume the usual factorization
\begin{equation}\label{spsdps}
\sigma_{\rm DPS}^{\psi \psi}=\frac{\sigma_{\rm SPS}^\psi \sigma_{\rm SPS}^\psi}{2\sigma_{\rm eff}}\;,
\end{equation}
 see~\cite{Bartalini:2011jp} for a general review. Here the factor of `2' is due to the identity of the final--state $J/\psi$ mesons, $\sigma_{\rm SPS}^\psi$ is the single $J/\psi$ photoproduction cross section and $\sigma_{\rm eff}$ contains information about the relative spatial distribution of the partons within the protons: the larger the average transverse distance between the partons, the larger $\sigma_{\rm eff}$ and the smaller the DPS cross section. For inclusive processes $\sigma_{\rm eff}$  is of order the proton radius squared $\sim 1$ ${\rm fm}^2$. However, in the case of photoproduction, due to the peripheral nature of this electromagnetic interaction, the average distance between partons in the colliding protons, and hence the corresponding $\sigma_{\rm eff}$ parameter, is much larger. To see this we note that the distribution in the squared momentum transfer in the photon--proton reaction is strongly peaked towards the kinematic minimum $t_{\rm min}$, which is given by (see e.g.~\cite{Jones:2013pga}) 
\begin{equation}
\sqrt{-t_{\rm min}}\approx \frac{M_\psi m_p}{\sqrt{s}}e^{\pm Y_\psi}\;.
\end{equation}
Considering production at LHCb, we can neglect the $-Y_X$ solution, which has a much smaller cross section (and in any case will have an even larger corresponding $\sigma_{\rm eff}$), and take $Y_X \sim 3$ as a representative value, to give a transverse separation of order $\sim 1/\sqrt{-t_{\rm min}} \sim 25$ fm, and so $\sigma_{\rm eff} \sim 625$ ${\rm fm}^2$, two to three orders of magnitude bigger than the inclusive value. Taking as an estimate the measured LHCb single $J/\psi$ cross photoproduction section~\cite{Aaij:2014iea}, we have $\sigma_{\rm SPS}^\psi \sim 5$ nb and so we from (\ref{spsdps}) we expect that 
\begin{equation}
 \sigma_{\rm DPS}^{\psi \psi} = O(10^{-2}) \,{\rm fb}\;.
\end{equation}
As we will see later, the expected cross section for the $gg \to J/\psi J/\psi$ SPS process is $O({\rm pb})$ or more, and so this DPS contribution can clearly be neglected.

\begin{figure}
\begin{center}
\includegraphics[scale=0.65]{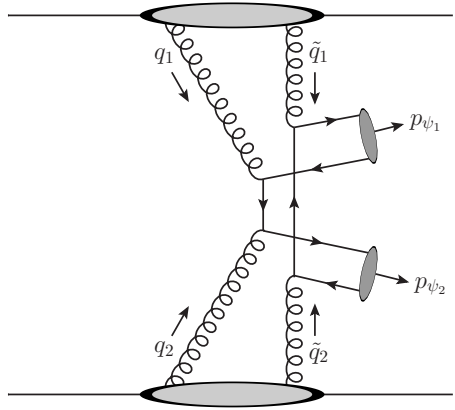}
\caption{Representative diagram for `symmetric' double $J/\psi$ CEP.}\label{fors}
\end{center}
\end{figure}

On the face of it then, we do not need to worry about DPS in the exclusive channel. However this is not quite the case: as well as the standard `skewed' CEP diagram of the type shown in Fig.~\ref{fig:pCp}, where only one of the $t$--channel gluon exchanges takes part in the hard $J/\psi$ pair production process, we must also in principle consider the `symmetric' diagram in Fig.~\ref{fors}. Such an interaction does not violate $C$--parity conservation, as the $c$ and $\overline{c}$ quarks which form the outgoing $J/\psi$ mesons are produced in separate $gg \to c\overline{c}$ subprocesses. Although it cannot truly be considered as DPS, due to the fact that these collisions are not in fact independent, this is nonetheless a correction to the simple `SPS' framework that we must  consider.

This `symmetric' type of diagram was first discussed in the context of light meson pair ($\pi\pi$, $\eta\eta$...) CEP in~\cite{HarlandLang:2011qd,HarlandLang:2012qz}, where this contribution was found to be small, even for the case of flavour--non--singlet $\pi\pi$ production, where the `skewed' cross section is already dynamically suppressed. It was in particular found that the `symmetric' cross section is power--suppressed, with an additional hard gluon propagator in general required to produce the meson pair final--state. However, this argument rested crucially on the fact that, within the approach of~\cite{HarlandLang:2011qd,HarlandLang:2012qz}, which made use of the `hard exclusive' formalism~\cite{Brodsky:1981rp,Benayoun:1989ng} to model the meson pair production subprocess, the momentum fraction $x$ of the quarks within the parent meson are integrated over and can in general take any value from $0$ to $1$, see~\cite{HarlandLang:2011qd} for a detailed discussion. For the specific case of $x=1/2$, which applies in the non--relativistic limit we consider here, this argument breaks down and \emph{a priori} we will expect no such suppression for the case of the `symmetric' diagram\footnote{In particular, in the forward proton limit and in the non--relativisitic quarkonium approximation we must have $q_{1\perp}=-q_{2\perp}=-\tilde{q}_{1\perp}=\tilde{q}_{2\perp}$ (for exact proton kinematics these equalities will approximately hold).}; we must therefore calculate this `symmetric' contribution explicitly.

The symmetric CEP amplitude can be written in the form
\begin{equation}\label{bsym}
A_{\rm sym.}=\pi^2 \int \frac{d^2 q_{1\perp}}{ q_{1\perp}^4 q_{2\perp}^4 }\,\overline{\mathcal{M}}_{\rm sym.}\,f_g(x_1,\tilde{x}_1, Q_{1}^2,\mu_F^2;t_1)f_g(x_2,\tilde{x}_2,Q_{2}^2,\mu_F^2;t_2) \; ,
\end{equation}
where the notation follows from Fig.~\ref{fors}: the $x_i$ are the momentum fractions carried by the gluons, while the scales $Q_i$ are defined as in (\ref{bt}), with $Q_1^2=Q_2^2=q_{1\perp}^2$ in the forward proton limit. The subprocess amplitude is given by
\begin{equation}\label{msym}
\overline{\mathcal{M}}_{\rm sym.}=\frac{4}{M_X^4}\frac{1}{N_C^2-1}\delta^{ac}\delta^{bd}q_{1\perp}^\mu q_{2\perp}^\nu \tilde{q}_{1\perp}^\alpha \tilde{q}_{2\perp}^\beta V^{ab}_{\mu\nu}\tilde{V}^{cd}_{\alpha\beta}\;,
\end{equation}
where $V$, $\tilde{V}$ are the standard $gg \to q\overline{q}$ vertices (with the tilde indicating which $q\overline{q}$ pair in Fig.~\ref{fors} is considered), but with the appropriate spin and colour projections performed on the outgoing quark/anti--quarks for $J/\psi$ pair production. The $f_g$'s in (\ref{bsym}) are the same generalized PDFs discussed in Section~\ref{durt}, but in the `time--like' domain, where the gluon momentum fractions have the same sign, see~\cite{HarlandLang:2011qd} for further discussion. In this region, the $f_g$'s cannot be extracted from DIS data, and the best we can do is to put an upper limit on the skewed PDFs by making use of the Schwarz inequality~\cite{Martin:1997wy}. This leads us to take

\begin{align}\nonumber
f_g(x,\tilde{x},Q^2,\mu_1^2,\mu_2^2) &=
\frac{1}{2}\frac{\partial}{\partial \log Q^2}\big[\sqrt{T_g(Q^2,\mu_1^2)T_g(Q^2,\mu_2^2)}
(xg(x,Q^2)+\tilde{x}g(\tilde{x},Q^2))\big]\;,\\ \label{fgsym}
& =
\frac{\partial}{\partial \log Q^2}\big[T_g(Q^2,\mu_F^2/4)
xg(x,Q^2)\big]\;,
\end{align}
where in the last line we have made use of the kinematic constraint on the gluon momentum fractions $x_{1,2}=\tilde{x}_{1,2}$, and the factorization scale $\mu_F$ is the same as in (\ref{fgskew}). We can see that there is no longer a square root on the Sudakov factor $T_g$, as both $t$--channel gluon exchanges take part in the interaction, but at a lower scale $\mu_F/2$, as each gluon pair that produces the $q\overline{q}$ has half the invariant mass of the $J/\psi J/\psi$ pair. Numerically, it is found that these PDFs lead to a factor of $\sim 2-3$ suppression relative to (\ref{fgskew}) in the relevant kinematic region. Although slightly different in form, we still find that (\ref{fgsym}) gives a gluon transverse momentum $Q_\perp^2$ of order a few ${\rm GeV}^2$, safely in the perturbative regime.

Making use of (\ref{fgsym}) and (\ref{bsym}) we can then readily calculate the predicted symmetric cross section. However, without performing the calculation explicitly, we in fact expect this symmetric contribution to be suppressed from general considerations. To see this, we can expand (\ref{msym}) as in (\ref{Agen}) in terms of the $g(\lambda_1)g(\lambda_2) \to q\overline{q}$  helicity amplitudes, $T_{\lambda_1\lambda_2}$, $\tilde{T}_{\lambda_1\lambda_2}$, as above with the appropriate spin and colour projections performed on the outgoing quark/anti--quarks. In the forward proton limit, we find that only a subset of the helicity amplitudes contribute, with
\begin{equation}\label{Agens}
\overline{\mathcal{M}}_{\rm sym.} \propto {\bf Q}_\perp^4((T_{++}+T_{--})(\tilde{T}_{++}+\tilde{T}_{--})+T_{+-}\tilde{T}_{-+}+T_{-+}\tilde{T}_{+-})\;.
\end{equation}
All other combinations (e.g. $T_{++}\tilde{T}_{+-}$...) are found to vanish in this forward limit (in some cases only after the $Q_\perp$ angular integration). Thus, while there is no individual requirement that the fusing gluons in the $gg \to q\overline{q}$ subprocess must be in a $J_z^P=0^+$ state, we can see from (\ref{Agens}) that the four gluon system (and hence the centrally produced $J/\psi$ pair) is still required to have these quantum numbers: either the fusing gluon pairs must be individually in a $J_z=0$ state, or with opposite $J_z=\pm 2$. Upon inspection we can also see that (\ref{Agens}) is symmetric under a parity transformation. The requirement that the centrally produced object must have $J_z=0$ quantum numbers in the forward limit can be derived purely from helicity conservation, independent of the details of the production process, see~\cite{Harland-Lang:2014lxa}, and so it is not surprising that we find this. As in the case of the standard `skewed' CEP mechanism, the parity constraint is on the other hand a model--dependent prediction of this perturbative approach.

If we now consider the threshold limit, then the $gg \to q\overline{q}$ amplitude for $|J_z|=2$ gluons will vanish due to conservation of angular momentum, and we are left with the even parity combinations of $J_z=0$ amplitudes in (\ref{Agens}). We have 
\begin{equation}\label{qqexc0}
T\left((g(\pm)g(\pm)\to q_h \overline{q}_{\bar{h}}\right) =\frac{16\pi\alpha_s}{N_c}\frac{m_q}{M_X}(\beta h \pm 1)\left(\frac{t^at^b}{1-\beta\cos\theta}+\frac{t^bt^a}{1+\beta\cos\theta}\right)\delta_{h,\bar{h}}\;.
\end{equation}
At threshold ($\beta=0$), the $g(\pm)g(\pm)\to q\overline{q}$ helicity amplitudes are therefore proportional to $\pm 1$ and so the even parity combination in (\ref{Agens}) will completely cancel. This is to be expected: as the $q\overline{q}$ system has parity $(-1)^{L+1}$, where $L$ is the orbital angular momentum, at threshold we have $L=0$ and the system is therefore in a $P=-1$ state. Thus we arrive at the result that the $J/\psi$ pair production process, via this symmetric mechanism, will completely vanish at threshold. More precisely, the decomposition (\ref{Agens}) is only valid up to to $O(Q_\perp^2/M_X^2)$ corrections due to the off--shellness of the fusing gluons, and these will not in general vanish at threshold. Nonetheless, as these are small corrections (although not completely negligible in the lower mass region\footnote{Strictly speaking such corrections correspond to higher--order QCD effects, which can generate this gluon off--shellness. While the $k_\perp$ factorization approach we use includes at LO part of this initial gluon off--shellness via the $q_{i\perp}$ dependent PDFs (\ref{fgskew}), due to the vanishing of the symmetric cross section at threshold in the on--shell limit (\ref{Agens}), the full cross section in this region will be strongly sensitive to precisely how such off--shell corrections are included and consequently to higher--order effects. Any calculation according to (\ref{bsym}) can therefore only be taken as an estimate, although even taking such an estimate in the lower mass region, a strong suppression relative the symmetric contribution is clear.}) we will still expect a strong suppression in the symmetric cross section in the threshold region. Finally, in the high--energy limit the $J_z=0$ contributions to (\ref{Agens}) will vanish (recalling that the $gg \to q\overline{q}$ amplitudes for massless quarks and $J_z=0$ gluons vanish), and only the individual $J_z=\pm 2$ terms remains. It can be confirmed analytically that the symmetric cross section is numerically suppressed in this regime, relative to the skewed cross section, although for brevity we do not show this explicitly here.

Finally, making use of (\ref{bsym}) we can give an estimated upper limit on the symmetric contribution to the double $J/\psi$ CEP cross section. Considering the $2<Y_{X}<4.5$ rapidity region relevant to the LHCb measurement discussed in the following section, and integrating from threshold we find, after an explicit calculation, that
\begin{equation}
 \frac{\sigma^{\rm sym.}}{\sigma^{\rm skew.}}\lesssim 3-5\,\%\;,
\end{equation}
with the precise ratio depending on the PDF choice and c.m.s. energy $\sqrt{s}$. We emphasise that this is an upper limit on the symmetric contribution, due to the fact that (\ref{fgsym}) corresponds to a maximum value of the generalized PDF, which may be smaller than this. Thus we can see that this `symmetric' mechanism is expected to give a very small contribution to the cross section, confirming the qualitative discussion above; in the following section, we will therefore safely consider just the pure `skewed' CEP cross section.

\section{Results}\label{res}

As discussed in the introduction, LHCb have recently reported the observation of exclusive $J/\psi J/\psi$ and $J/\psi \psi(2S)$ production~\cite{Aaij:2014rms} at $\sqrt{s}=7$ and 8 TeV. After correcting for a proton dissociative background, this corresponds to a cross section of
\begin{equation}\label{lhcbdat}
 \sigma^{J/\psi J/\psi} = 24 \pm 9 \,{\rm pb}\;,
\end{equation}
with
\begin{equation}\label{lhcbdatr}
\frac{\sigma(J/\psi \psi(2S))}{\sigma(J/\psi J/\psi)}=1.1^{+0.5}_{-0.4}\;,
\end{equation}
where in both cases the cross sections correspond to the rapidity regions $2<Y_{X}<4.5$, and the latter results assumes the same elastic fraction for both processes. This constitutes the only measurement of exclusive double charmonia production, and there are encouraging possibilities in the future for higher statistics measurements to be made~\cite{Ronanpriv}. We therefore concentrate in this paper on this kinematic region, although we will also show results for production at central rapidities. We take the pure non--relativistic limit throughout with $m_c=M_\psi/2$, $\mu_F=\mu_R=m_\perp=\sqrt{M_\psi^2+p_{\psi,\perp}^2}$, as in~\cite{Qiao:2009kg,Berezhnoy:2011xy} and the LO expression for $\alpha_s(\mu_R)$, which is used in~\cite{Berezhnoy:2011xy}, where a good agreement with the inclusive $J/\psi J/\psi$ data is found, see also the discussion below. We fix the value of the $J/\psi$ wave function at the origin to its leptonic width~\cite{Beringer:1900zz}, with
\begin{equation}\label{r0val}
 |R_0(0)|^2=0.56 \, {\rm GeV}^3\;.
\end{equation}
We note that often somewhat larger values ($\sim 0.9\,{\rm GeV}^3$) for this parameter are taken in the literature, see for example~\cite{Qiao:2009kg,Kom:2011bd}, however such a higher value is typically a result of fits to inclusive charmonia data that include relativistic as well as higher--order in $\alpha_s$ corrections to the calculations, see~\cite{Bodwin:2007fz}. We take the lower value (\ref{r0val}), which is extracted without including such corrections, as these are not included in our calculation, and we would therefore argue this is the more consistent choice to take. In~\cite{Hoodbhoy:1996zg}, it was moreover shown that relativistic and higher--order corrections are numerically small and partly cancel each other in the case of $J/\psi$ photoproduction, if such a normalization is taken. This choice is also supported by the fact that in for example~\cite{Berezhnoy:2011xy}, such a value was taken, and good agreement with the LHCb inclusive double $J/\psi$ data~\cite{Aaij:2011yc} was found, within experimental and theoretical uncertainties, when similar LO PDFs to those we take here are used\footnote{In~\cite{Kom:2011bd} good agreement with the LHCb data is also found, using the much higher value as in~\cite{Bodwin:2007fz}, however here NLO PDFs are used, which are smaller in the low $x$ and $Q^2$ region relevant to this process, and this choice largely cancels the effect of the increased $|R_0(0)|^2$.}. Moreover, in the analysis~\cite{Jones:2013pga} of exclusive $J/\psi$ photoproduction, a good description of the available data is found with this choice of $|R_0(0)|^2$. However, if the higher value of $|R_0(0)|^2$ is used, the resultant double $J/\psi$ cross section, which depends on this parameter squared, is a factor of $\sim 3$ larger. While, as discussed above, we do not find that this is a suitable choice for the calculation we are considering, this is nevertheless an indication of the fairly large theoretical uncertainty that arises due to the value of $|R_0(0)|^2$ and to the related issue of the importance of (process--dependent) relativistic corrections. 

\begin{table}
\begin{center}
\begin{tabular}{|l|c|c|c|c|}
\hline
&model 1&model 2&model 3&model 4\\
\hline
MSTW08LO &2.3&5.2&3.3&2.7\\
\hline
CTEQ6L &1.4&3.1&2.0&1.6\\
 \hline
 GJR08LO &2.6&5.7&3.7&3.0\\
 \hline
\end{tabular}
\caption{Cross section, in pb, for the $\gamma\gamma$ CEP at $\sqrt{s}=1.96$ TeV, for representative choice of LO PDFs and models of the soft survival factor. The photons are required to have $E_\perp^\gamma>2.5$ GeV and $|\eta^\gamma|<1$.}\label{psigam}
\end{center}
\end{table}

There is in general quite a large uncertainty in the predictions for CEP due to the choice of PDF and model of the soft survival factor, and so here we will use the CDF measurement of exclusive $\gamma\gamma$ production~\cite{Aaltonen:2011hi} at $\sqrt{s}=1.96$ TeV as guidance, for which a sample of 43 dominantly exclusive $\gamma\gamma$ events were measured, with $E_\perp^\gamma>2.5$ GeV and $|\eta^\gamma|<1$, corresponding to a cross section of
\begin{equation}\nonumber
 \sigma^{\gamma\gamma}=2.48^{+0.40}_{-0.35}\,({\rm stat})\,{}^{+0.40}_{-0.51}\,({\rm syst}) \, {\rm pb}\;.
\end{equation}
Up--to--date predictions for this event selection are shown in Table~\ref{psigam}, and we can see that for a range of representative LO PDFs and soft survival models there is good agreement with the CDF measurement\footnote{These predictions are different from those presented previously in~\cite{HarlandLang:2012qz}: this is due to the fact we now use the updated prescription discussed in~\cite{Harland-Lang:2013xba} to calculate the skewed PDFs and in addition now consistently take the LO expression for $\alpha_s$ from each of the corresponding PDF sets.},  with the MSTW08LO~\cite{Martin:2009iq} (models 1 and 4), CTEQ6L (LO $\alpha_s$)~\cite{Pumplin:2002vw} (models 2 and 3) and GJR08LO (fixed flavour)~\cite{Gluck:2007ck} (models 1 and 4) lying in the preferred 2--3 pb range. Given this agreement, we can use these choices to improve the reliability of our predictions for double $J/\psi$ production at the LHC. However, we note that the CDF measurement (for which the photons are produced centrally, with $|\eta_\gamma|<1$) corresponds to a quite different $x$ region, with $x \sim M_{\gamma\gamma}/\sqrt{s} \sim 3 \times 10^{-3}$, to that probed at the LHC for double $J\/psi$ production: in the forward region relevant to the LHCb measurement, at the higher LHC energies, this spans a large range of $x$, from $\sim 10^{-5}$ -- $10^{-1}$ (recall that the uncertainty in the gluon PDF at low $x$ and $Q^2$ is in particular quite large). In addition, we are performing such a comparison purely at LO, whereas in general we might expect (process--dependent) higher--order corrections to affect the results of such a comparison, and there is also an important uncertainty in size of the survival factors, in particular in the energy dependence in going from Tevatron to LHC energies. Thus this comparison with the CDF data can only be considered as a guideline.

\begin{table}
\begin{center}
\begin{tabular}{l|c|c|c|c|c|c|}
\cline{2-7}
&\multicolumn{2}{c|}{MSTW08LO}&\multicolumn{2}{|c|}{CTEQ6L}&\multicolumn{2}{|c|}{GJR08LO}\\
\cline{2-7}
  &model 1 &model 4&model 2&model 3&model 1& model 4\\
\hline
\multicolumn{1}{|l|}{$\sqrt{s}=8$ TeV} &5.8&6.7&2.5&1.9&4.6&5.4\\
\hline
\multicolumn{1}{|l|}{$\sqrt{s}=14$ TeV} &12&14&4.7&3.6&9.1&11\\
\hline
\end{tabular}
\caption{Cross section, in pb, for the $J/\psi J/\psi$ CEP at $\sqrt{s}=8$ and 14 TeV, using different models of the soft survival factor taken from~\cite{Khoze:2013dha}, and with different PDF choices. The $J/\psi$ pair is required to lie in the the forward rapidity region $2<Y_X<4.5$.}\label{psics}
\end{center}
\end{table}

\begin{table}
\begin{center}
\begin{tabular}{l|c|c|c|c|c|c|}
\cline{2-7}
&\multicolumn{2}{c|}{MSTW08LO}&\multicolumn{2}{|c|}{CTEQ6L}&\multicolumn{2}{|c|}{GJR08LO}\\
\cline{2-7}
  &model 1 &model 4&model 2&model 3&model 1& model 4\\
\hline
\multicolumn{1}{|l|}{$\sqrt{s}=8$ TeV} &15&17&10&8&16&19\\
\hline
\multicolumn{1}{|l|}{$\sqrt{s}=14$ TeV} &25&31&17&13&30&36\\
\hline
\end{tabular}
\caption{Cross section, in pb, for the $J/\psi J/\psi$ CEP at $\sqrt{s}=8$ and 14 TeV, using different models of the soft survival factor taken from~\cite{Khoze:2013dha}, and with different PDF choices. The $J/\psi$ pair is required to lie in the the central rapidity region $-2.5<Y_X<2.5$.}\label{psicscent}
\end{center}
\end{table}

\begin{figure}
\begin{center}
\includegraphics[scale=0.6]{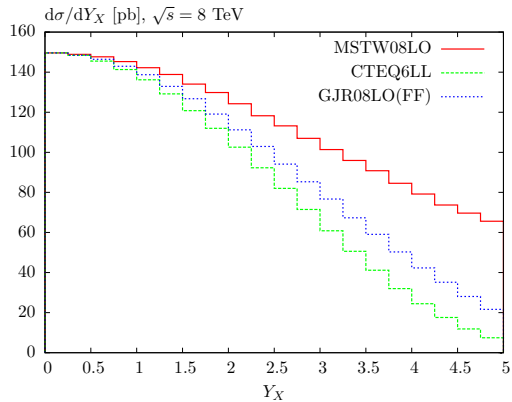}
\caption{Distribution in rapidity of the central system, $Y_X$, for $J/\psi J/\psi$ CEP at $\sqrt{s}=8$ TeV and for different choices of PDF. For display purposes, the first bins are normalized to the MSTW08LO result.}\label{psirap}
\end{center}
\end{figure}

In Table~\ref{psics} we show predictions for the LHCb acceptance ($2<Y_{X}<4.5$) at $\sqrt{s}=8$ and 14 TeV\footnote{The LHCb measurements in fact correspond to data taken at both $\sqrt{s}=7$ and 8 TeV: we show predictions for the latter energy for simplicity, but note that there is only a small difference between the predictions at these two energies.}, using the PDF sets and model choices described above. These predictions are calculated using a full MC simulation of the CEP process, including the $J/\psi \to \mu^+\mu^-$ decay with spin correlations, and will be released as part of a publicly available \texttt{SuperCHIC 2} MC in the near future~\cite{HLfut}. The results in Tables~\ref{psics} and~\ref{psicscent} are presented with the rapidity cut placed on the double $J/\psi$ system and not the decay products, to be consistent with the LHCb Run--I result\footnote{In~\cite{Aaij:2014rms} the quoted cross section for $2<Y_X<4.5$ results from a correction to the data which assumes a particular rapidity and mass distribution of the double $J/\psi$ system $X$ and that the $J/\psi \to \mu^+\mu^-$ decay occurs isotropically. Although in the high subprocess energy limit, we expect the $J/\psi$ mesons to be dominantly transversely polarized, see e.g. Fig.~\ref{ang}, we find that for the total cross section integrated from threshold, the $J/\psi$ mesons can in fact to reasonable approximation be treated as unpolarized.}. We can see that the MSTW08LO and GJR08LO PDFs give a factor of $\sim 2$ -- 3 larger cross sections than the CTEQ6L set, despite the agreement between sets found in Table~\ref{psigam} for $\gamma\gamma$ CEP at $\sqrt{s}=1.96$ TeV, which is precisely due to the different $x$ values probed in the current case, and the different forms of the PDFs in these regions: the CTEQ6L PDFs in particular give a cross section that falls off more sharply at higher rapidity. This is seen in Fig.~\ref{psirap}, where the rapidity distribution of the central system is shown for the choices of PDF discussed above (it is worth pointing out that the shape of the rapidity distribution is sensitive not only to the input PDF but also the skewedness which is included as in~\cite{Shuvaev:1999ce,Harland-Lang:2013xba}). The range of predictions in Table~\ref{psics} therefore gives some estimate of the uncertainty in our predictions due the choice of PDF and model of soft survival factor, although as discussed above, due to the different $x$ values probed and higher c.m.s. energy this can only be taken as a guide. We must also consider the possibility of higher--order corrections: varying the renormalization scale $\mu_R$ of $\alpha_s$ in the subprocess matrix element between $m_\perp/2$ and $2m_\perp$ gives a large variation of order $\sim{}^{\times}_{\div} 2-3$, due to the high power of $\alpha_s$ in the calculation (this uncertainty is not limited to the exclusive case: a similar level of variation is seen in the inclusive process, when the factorization and renormalization scales are allowed to vary independently~\cite{Lansberg:2013qka}), and is even larger if the factorization scale is varied independently as well. This is clearly a significant source of uncertainty, although we can take as guidance the fact that the default choice of $\mu_F=\mu_R=m_\perp$ gives a good description of the existing data in the inclusive case~\cite{Berezhnoy:2011xy}. Finally, we note that relativistic corrections have not been included in our calculation. It is worth recalling, as discussed in Section~\ref{sec:thres} that the $gg \to J/\psi J/\psi$ amplitudes with $J_z=0$ incoming gluons are numerically suppressed near threshold relative to the amplitudes with $|J_z|=2$ incoming gluons, while they vanish entirely in the high $M_X$ limit: this leads the predicted CEP cross section to be correspondingly reduced beyond naive estimations. If relativistic corrections counteract this suppression at all, the predicted CEP cross section could also be larger.

\begin{figure}
\begin{center}
\includegraphics[scale=0.8]{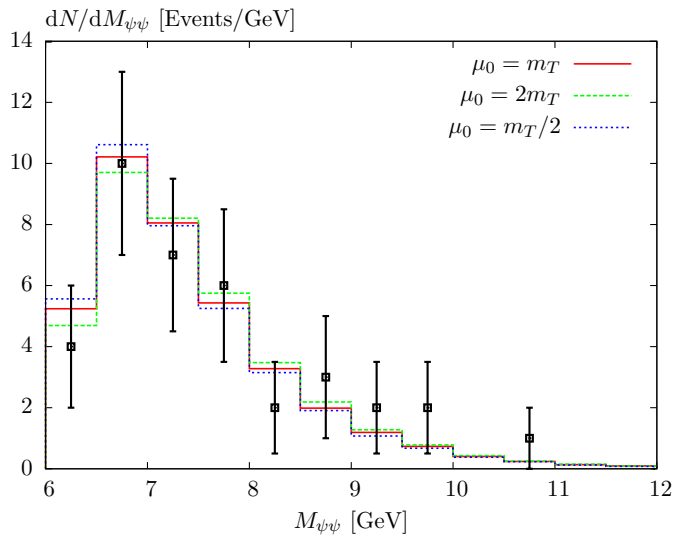}
\caption{Comparison of LHCb measurement~\cite{Aaij:2014rms} of $J/\psi J/\psi$ invariant mass distribution with theory prediction, calculated as described in the text. In all cases the result is normalized to the data.}\label{lhcbm}
\end{center}
\end{figure}

Given this discussion above, we can see from Table~\ref{psics} that there is fair agreement with the LHCb measurement (\ref{lhcbdat}), within the fairly large theoretical and experimental uncertainties, and with the higher predictions from the MSTW08LO and GJR08LO PDF predictions being favoured. Moreover, we show in Fig.~\ref{lhcbm} a comparison of the LHCb measurement~\cite{Aaij:2014rms} of the $J/\psi J/\psi$ invariant mass distribution with our prediction, made using MSTW08LO PDFs and normalized to the data. We can see that the shape of the distribution is very well described, within experimental uncertainties (we recall that in the case of inclusive double $J/\psi$ production, there are indications of some discrepancy between the predicted mass distributions and the data~\cite{Aaij:2011yc}). As described above, the theoretical curves are calculated using a full MC simulation of the double $J/\psi \to \mu^+\mu^-$ CEP process, including spin correlations in the $J/\psi$ decay, with the final--state muons required to have $2<\eta_\mu<4.5$. To give some estimate of the theoretical uncertainty on this distribution, we also show predictions corresponding to varying the renormalization and factorization scales by a factor of 2 up and down, and we can see that the shape of the distribution is relatively insensitive to this (there is some variation in the shape due to the PDF choice, although this is also small). Thus the theoretical uncertainty on the shape of this distribution, which is driven by the form of the contributing matrix elements, as well as factors specific to the exclusive channel, such as the $M_X$ dependent Sudakov factor in (\ref{fgskew}), is much smaller than that in the total cross section normalization (we recall this scale variation gives a $\sim {}^{\times}_{\div}$ 3 spread in the cross section). Clearly a higher statistics measurement of exclusive double $J/\psi$ production, which would allow a closer comparison between theory and data, both in the absolute cross section normalization and invariant mass (and other) distributions, is desirable. We can see from Table~\ref{psics} that the predicted cross sections for $\sqrt{s}=14$ TeV are a factor of $\sim 2$ larger, while in Table~\ref{psicscent} we show predictions for central rapidities at both $\sqrt{s}=8$ and 14 TeV, and can see that the predicted cross sections can be larger still.

\begin{figure}
\begin{center}
\includegraphics[scale=0.7]{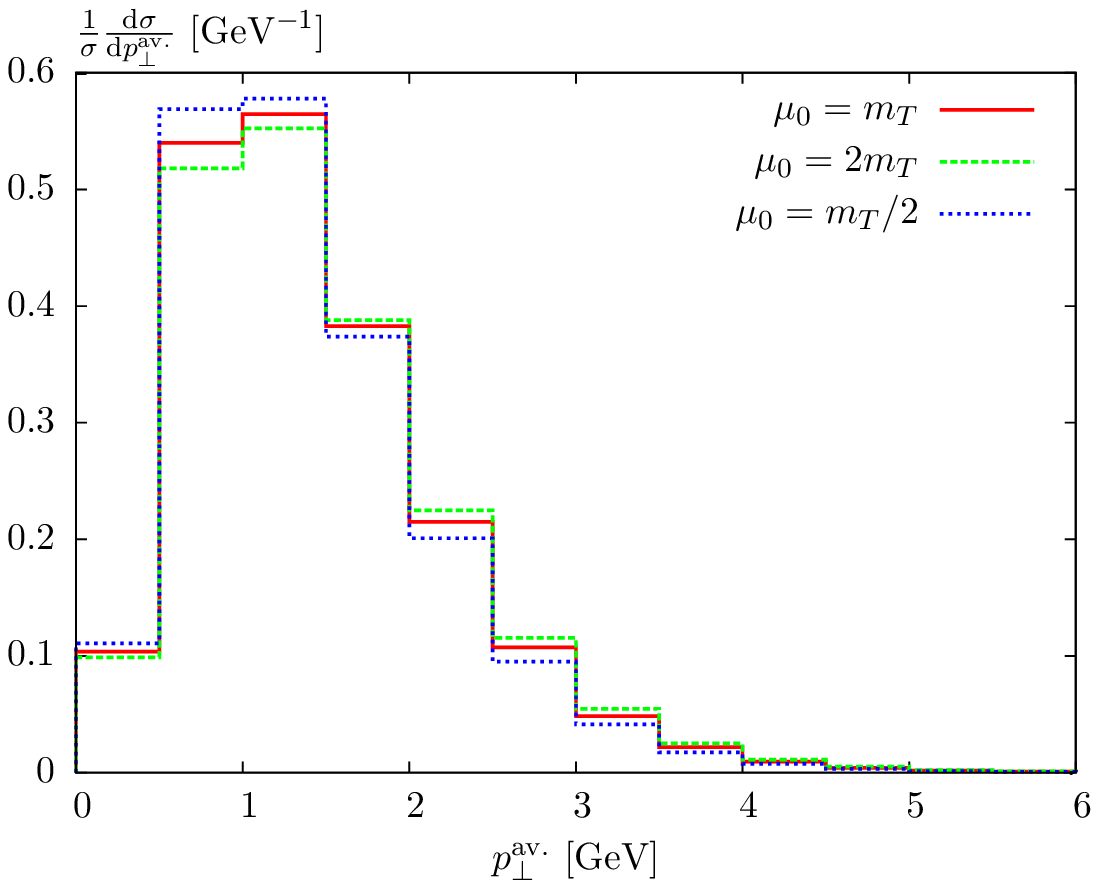}
\includegraphics[scale=0.7]{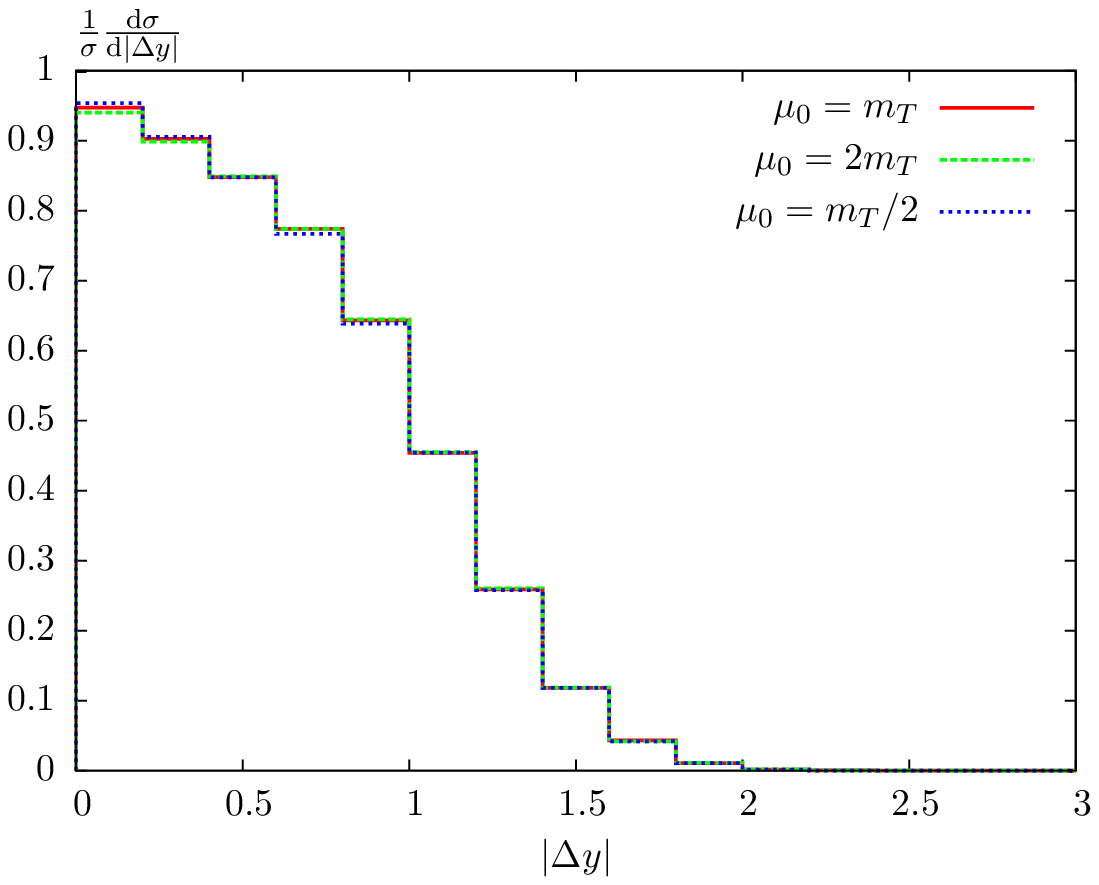}
\caption{(Left) Normalized distribution with respect to the scalar average transverse momentum $p_\perp^{\rm av.}=(|p_\perp^{\psi_1}|+|p_\perp^{\psi_2}|)/2$. (Right) Normalized distribution with respect to the rapidity difference $|\Delta y|=|y_1^{\psi_1}-y_2^{\psi_2}|$. In both cases distributions for different choices of factorisation and renormalisation scale are shown.}\label{ptdy}
\end{center}
\end{figure}

It is clear from the discussion above that there are some important uncertainties in the absolute cross section predictions for $J/\psi J/\psi$ CEP. One possibility to reduce these is to consider more differential observables, the shape of which will be much less sensitive to these uncertainties. An example of this is the $J/\psi$ pair invariant mass distribution shown in Fig.~\ref{lhcbm} and discussed above. It is also interesting to consider the $J/\psi$ transverse momentum distribution, as shown in Fig.~\ref{ptdy} (left): here the scalar average of the $J/\psi$ transverse momenta is taken (the individual transverse momenta are in general not exactly equal due to the non--zero proton $p_\perp$ in the final state), although other choices of suitable variable are possible. In addition, we can consider the rapidity separation, $\Delta y$, between the $J/\psi$ mesons. We recall that this variable is of much interest in the inclusive case, where the DPS distribution is expected to much broader than the SPS one~\cite{Kom:2011bd,Lansberg:2014swa}. Exclusively, we have found that the DPS contribution is expected to be very small, however it is interesting to observe that the $\Delta y$ distribution from the exclusive SPS contribution is also broader than the corresponding inclusive SPS contribution (see for example Fig.~4 of~\cite{Kom:2011bd}). This is in fact due to the selection rule which operates for CEP: the amplitudes for which the initial--state gluons are in a $J_z=0$ state are dynamically enhanced, and these are found to be strongly peaked towards $|\cos \theta| \to 1$ (where $\theta$ is the scattering angle in the $J/\psi$ pair rest frame), while wide angle scattering, $\cos \theta \sim 0$, is suppressed, see for example the dashed curve in Fig.~\ref{ang}. This results in a strong suppression in the low $|\Delta y|$, wide--angle scattering region, while the full distribution, which is expected to be sensitive to both $J_z=0$ and $|J_z|=2$ configurations in the lower mass region (see (\ref{supthres}) and the discussion below) is a non--trivial combination of both contributions. A measurement of this rapidity difference, which is sensitive to the relative combination of these helicity configurations, would therefore be of great interest as a probe of the underlying CEP selection rule. 

A further possibility to reduce the theoretical uncertainties is to consider ratios of observables: in Table~\ref{psirats} we show the ratio of the double $J/\psi$ to the $\gamma\gamma$ and $\phi\phi$ CEP cross sections (the latter calculated using the formalism described in~\cite{HarlandLang:2011qd}, with a choice of distribution amplitude described in Appendix~\ref{app:mes}), within the same invariant mass regions, with the photon and $\phi$ transverse momentum required to have $p_\perp>2.5$ GeV in order to ensure perturbatively reliable cross sections. These ratios are almost independent of the model of soft survival and PDF used, and consequently have much smaller theoretical uncertainties than for the absolute cross sections considered above. We emphasise that the double $J/\psi$ CEP cross section is suppressed beyond naive estimations due the numerical suppression in the $gg \to J/\psi J/\psi$ amplitudes, while in the $\gamma\gamma$ case there is no such suppression, and the predicted $\phi\phi$ cross section,  which is also found to be dynamically suppressed, results from the non--trivial calculation described in~\cite{HarlandLang:2011qd}. A measurement of these ratios would therefore represent a non--trivial test of the underlying theory.

\begin{table}[h]
\begin{center}
\begin{tabular}{|l|c|c|c|}
\hline
&$\frac{\sigma(J/\psi J/\psi)}{\sigma(\gamma\gamma)}$&$\frac{\sigma(J/\psi J/\psi)}{\sigma(\phi\phi)}$&$\frac{\sigma(J/\psi J/\psi)}{\sigma(\pi^+\pi^-)}$\\
\hline
$M_X>2M_{\psi}$ &0.76&5.5&37\\
$M_X>5$ GeV &0.49&3.9&11\\
 \hline
\end{tabular}
\caption{Ratio of $J/\psi J/\psi$ to $\gamma\gamma$ and $\phi\phi$ CEP cross section at $\sqrt{s}=8$ TeV. The central system is required to lie in the the rapidity region $2<Y_X<4.5$, and the photons and $\phi$, $\pi$ mesons are required to have transverse momentum $p_\perp>2.5$ GeV. Results are also shown for a lower cut on the central system invariant mass $M_X>5$ GeV, in the case of $\gamma\gamma$, $\phi\phi$ and $\pi^+\pi^-$ production.}\label{psirats}
\end{center}
\end{table}

It is also of interest to consider the ratios of double $\psi(2S)$ and $J/\psi \psi(2S)$ CEP to double $J/\psi$ CEP, in particular in light of the LHCb measurement (\ref{lhcbdatr}) discussed above. While a complete calculation in the $\psi(2S)$ must involve a proper treatment of relativistic effects, to give an estimate we normalize to the $\psi(2S)$ leptonic width to get
\begin{equation}
 |R^{\psi(2S)}(0)|^2=0.43\, {\rm GeV}^3\;.
\end{equation}
and, as in~\cite{Berezhnoy:2011xy}, we simply take $m_c=(M_\psi+M_{\psi^\prime})/4$ in the $J/\psi \psi(2S)$ case and $m_c=M_{\psi^\prime}/2$ in the double $\psi(2S)$ case. Using this, we find that
\begin{equation}
 \sigma^{J/\psi J/\psi} \, :\, \sigma^{J/\psi \psi(2S)}\, : \, \sigma^{\psi(2S) \psi(2S)} = 1\,:\,0.40\,:\,0.044\;,
\end{equation}
for the LHCb acceptance at $\sqrt{s}=8$ TeV, integrating from threshold. This ratio is largely independent of the rapidity region considered as well as the PDF choice and model of soft survival factor. We can see that the predicted  $J/\psi \psi(2S)$ ratio is in reasonable agreement with the measured ratio (\ref{lhcbdatr}), within the large experimental uncertainties, although perhaps somewhat lower. Clearly with further higher statistics data (noting that the systematic uncertainties in the measured ratio largely cancel~\cite{Aaij:2014rms}) a better comparison can be made, including potentially the double $\psi(2S)$ case, which has not been seen so far by LHCb.

It is also interesting to consider the possibility for observing the exclusive pair production of $C$--even $\chi_{cJ}$ states. These were also searched for but not seen by LHCb~\cite{Aaij:2014rms}, via the $\chi_c \to J/\psi \gamma$ decay chain, with limits set on the cross sections in the $\chi_{c1}$ and $\chi_{c2}$ cases of the same order as the observed double $J/\psi$ cross section (in the $\chi_{c0}$ case the limit was much higher due to the lower branching to $J/\psi \gamma$). This is not surprising: due to the much lower formation probability in the $\chi_c$ case the predicted CEP cross section is expected to be significantly lower. Using the standard expressions for the $J/\psi \to ggg$ and $\chi_{c0} \to gg$ decay widths we have
\begin{equation}\label{chic}
 \frac{|R_1^\prime(0)|^2}{m_c^2 |R_0(0)|^2} = \frac{5\alpha_s}{18} \frac{2(\pi^2-9)}{27\pi} \frac{\Gamma_{\chi_0}}{\Gamma_\psi}\;\approx 0.6 \,\alpha_s\;,
\end{equation}
where $R_1^\prime(0)$ is the derivative of the $P$--wave wave function at the origin, and we have divided by $m_c^2$ as the $P$--wave formation amplitude is proportional to $R_1^\prime(0)/m_c$. We also have assumed in the last step that these gluonic widths of the $\chi_c$ and $J/\psi$ mesons are given by the measured total widths~\cite{Beringer:1900zz}. As the ratio of the double $\chi_c$ to double $J/\psi$ cross sections is proportional to the square of (\ref{chic}), we therefore expect a roughly 2 orders of magnitude suppression in the double $\chi_c$ CEP cross section. However, we recall that the double $J/\psi$ CEP cross section is numerically suppressed, due to the structure of the $gg \to J/\psi J/\psi$ helicity amplitudes and the $J_z=0$ selection rule which operates in CEP: in the case of double $\chi_c$ production this may not be the case, and as these $C$--even states can couple individually to two gluons, the `ladder' type of diagrams discussed in~\cite{HarlandLang:2011qd,Harland-Lang:2013ncy} can contribute, and these may not be suppressed in this way. This may therefore compensate some of the effect of the smaller $\chi_c$ formation probability. Finally, we note that the formation amplitude of the pseudoscalar $\eta_c$ meson is proportional to the same value of the wave function at the origin $R_0(0)$ as in the $J/\psi$ case, and may also be produced by this additional type of `ladder' diagram: we may therefore expect the cross section for exclusive $\eta_c \eta_c$ production to be of the same size or even bigger than the $J/\psi J/\psi$ cross section.

\section{Conclusion}\label{conc}

In this paper we have presented the first calculation of the exclusive production of charmonia ($J/\psi, \psi(2S)$) pairs in hadronic collisions. Such a process is of great interest as a probe of the underlying formalism in the exclusive calculation, and is sensitive to the direct (e.g. $gg \to J/\psi J/\psi$) channel, with contributions from double parton scattering events expected to be extremely small. By carefully considering the form of the  $gg \to J/\psi J/\psi$ helicity amplitudes, we have shown that we expect the CEP cross section for double $J/\psi$ to be suppressed below naive expectations, due to the dynamical $J^P_z=0^+$ selection rule which operates. We have then seen that the non--trivial predictions `Durham' pQCD--based model of CEP are in fair agreement with the recent LHCb measurement of exclusive double $J/\psi$ production~\cite{Aaij:2014rms}, within theoretical and experimental uncertainties, although potentially underestimating the observed signal somewhat. Moreover, the measured invariant mass distribution is found to be well described by the theory, with no hint of the discrepancy which may be present in the inclusive case~\cite{Aaij:2011yc}.

While these results are encouraging, it is clear that further data on charmonia production from future LHC runs will be able to place a much tighter constraint on the theoretical predictions, as well as being more sensitive to possible exotic (tetraquark...) contributions. We have also discussed how measurements of ratios of observables can help reduce the theoretical uncertainties, and considered predictions for the higher energy LHC runs. The exclusive production of charmonia pairs opens up a rich field of studies: for example the possible importance of relativistic corrections has yet to be fully addressed, and the $\chi_{cJ}$ and $\eta_c$ charmonia states represent other potential observables not considered in detail here. The results discussed in this paper and the encouraging agreement with the first LHCb measurement of exclusive charmonia pair production provide strong motivation for such theoretical work and for further measurements in the future. 

\section*{Acknowledgements}
We thank Ronan McNulty for many useful discussions and Guy Wilkinson for encouragement in the preparation of this work. We also thank Anatoli Likhoded and Alexey Novoselov for some helpful correspondences. MGR thanks the IPPP for hospitality. This work was supported by the RSCF grant 14-22-00281.

\appendix

\section{Threshold limit: general forms}\label{app:thres}

The general form of the $gg \to J/\psi J/\psi$ amplitudes in the threshold limit, discussed in Section~\ref{sec:thres} are given by
\begin{align}\nonumber
T_{\lambda_1,\lambda_2,\lambda_3,\lambda_4}=&\frac{2\pi\alpha_s^2|R_0(0)|^2}{9M_\psi^3}\big(15(\lambda_1\lambda_2-1)+2+\lambda_3\lambda_4(1+\lambda_1\lambda_2) \\ \label{thres1}
&-14(\lambda_1-\lambda_2)(\lambda_3-\lambda_4)\cos\theta+14\lambda_3\lambda_4(1-\lambda_1\lambda_2)\cos^2\theta\big)\;,\\ \label{thres2}
T_{\lambda_1,\lambda_2,\lambda_3,0}=&\frac{56\pi\alpha_s^2|R_0(0)|^2}{9\sqrt{2}M_\psi^3}\sin\theta\big((\lambda_2-\lambda_1)-\lambda_3(1-\lambda_1\lambda_2)\cos\theta\big)\;,\\\label{thres3}
T_{\lambda_1,\lambda_2,0,\lambda_4}=&-\frac{56\pi\alpha_s^2|R_0(0)|^2}{9\sqrt{2}M_\psi^3}\sin\theta\big((\lambda_2-\lambda_1)+\lambda_4(1-\lambda_1\lambda_2)\cos\theta\big)\;,\\  \label{thres4}
T_{\lambda_1,\lambda_2,00}=&\frac{4\pi\alpha_s^2|R_0(0)|^2}{9M_\psi^3}\big(15-13\lambda_1\lambda_2-14(1-\lambda_1\lambda_2)\cos^2\theta\big)\;,
\end{align}
where $\lambda_i$ are the transverse helicities of the particles.

\section{$\phi$ meson distribution amplitude}\label{app:mes}

In general, within the `hard exclusive' formalism the meson distribution amplitudes can be expanded in terms of the Gegenbauer polynomials $C_n$~\cite{Lepage:1980fj,Baier:1981pm} (see also~\cite{Harland-Lang:2013ncy} for more discussion)
\begin{equation} \label{waves}
\phi_{M}(x,\mu_F^2)=\frac{6 f_M}{2\sqrt{N_C}} x(1-x)[1+\sum_{n=2,4,\cdots} a_n(\mu_F^2)C_n^{3/2}(2x-1)]\;,
\end{equation}
where $\mu_F$ is the factorization scale, taken as usual to be of the order of the hard scale of the process being considered, and $f_M$ is the meson decay constant: although for the case of the vector $\phi$ meson, this is in general a polarization dependent object, here we take a universal $f_\phi=230$ MeV, as in~\cite{Benayoun:1989ng}. The evolution of the distribution amplitude is dictated by the $\mu_F^2$ dependence of the coefficients $a_n$, which evolve to 0 at asymptotically high energies. The higher--order $n=4,6,$... terms evolve faster towards zero with increasing $n$, and combined with the fact that as $n$ increases, the additional powers of $C_n^{3/2}(2x-1)$ give a smaller numerical contribution to the distribution amplitude, this means that any fit can effectively truncate the series (\ref{waves}) after a limited number of terms. For the case of the transversely polarized $\phi$ meson we truncate at $n=4$ and use the fit of~\cite{Benayoun:1989ng}, which extracts the expectation values
\begin{equation}
 \langle \xi^n \rangle = \int_{-1}^{1} \, {\rm d}\xi\,\xi^n\phi(x)\;,
\end{equation}
where $\xi=2x-1$, from QCD sum rules. This gives $a_2=-1/3$ and $a_4=11/80$. In the case of the longitudinal polarizations, the $\phi$ distribution amplitude is found to be approximately asymptotic~\cite{Benayoun:1989ng}, i.e. with $a_n=0$ for $n\geq 2$.

\bibliography{references}{}

\begin{thebibliography}{10}

\bibitem{Harland-Lang:2014lxa}
L.~Harland-Lang, V.~Khoze, M.~Ryskin, and W.~Stirling,
\newblock Int.J.Mod.Phys. {\bf A29}, 1430031 (2014), 1405.0018.

\bibitem{Harland-Lang:2014dta}
L.~Harland-Lang, V.~Khoze, and M.~Ryskin,
\newblock Int.J.Mod.Phys. {\bf A29}, 1446004 (2014).

\bibitem{Albrow:2010yb}
M.~G. Albrow, T.~D. Coughlin, and J.~R. Forshaw,
\newblock Prog.Part.Nucl.Phys. {\bf 65}, 149 (2010), 1006.1289.

\bibitem{Qiao:2009kg}
C.-F. Qiao, L.-P. Sun, and P.~Sun,
\newblock J.Phys. {\bf G37}, 075019 (2010), 0903.0954.

\bibitem{Kom:2011bd}
C.~Kom, A.~Kulesza, and W.~Stirling,
\newblock Phys.Rev.Lett. {\bf 107}, 082002 (2011), 1105.4186.

\bibitem{Novoselov:2011ff}
A.~Novoselov,
\newblock (2011), 1106.2184.

\bibitem{Baranov:2011ch}
S.~Baranov, A.~Snigirev, and N.~Zotov,
\newblock Phys.Lett. {\bf B705}, 116 (2011), 1105.6276.

\bibitem{Lansberg:2013qka}
J.-P. Lansberg and H.-S. Shao,
\newblock Phys.Rev.Lett. {\bf 111}, 122001 (2013), 1308.0474.

\bibitem{Lansberg:2014swa}
J.-P. Lansberg and H.-S. Shao,
\newblock (2014), 1410.8822.

\bibitem{Aaij:2011yc}
LHCb Collaboration, R.~Aaij {\em et~al.},
\newblock Phys.Lett. {\bf B707}, 52 (2012), 1109.0963.

\bibitem{CMS:2013pph}
CMS collaboration,
\newblock CMS-PAS-BPH-11-021.

\bibitem{Abazov:2014qba}
D0 Collaboration, V.~M. Abazov {\em et~al.},
\newblock (2014), 1406.2380.

\bibitem{Berezhnoy:2011xy}
A.~Berezhnoy, A.~Likhoded, A.~Luchinsky, and A.~Novoselov,
\newblock Phys.Rev. {\bf D84}, 094023 (2011), 1101.5881.

\bibitem{Aaij:2014rms}
LHCb Collaboration, R.~Aaij {\em et~al.},
\newblock (2014), 1407.5973.

\bibitem{Khoze97}
V.~A. Khoze, A.~D. Martin, and M.~G. Ryskin,
\newblock Phys. Lett. {\bf B401}, 330 (1997), hep-ph/9701419.

\bibitem{Khoze00}
V.~A. Khoze, A.~D. Martin, and M.~G. Ryskin,
\newblock Eur. Phys. J. {\bf C14}, 525 (2000), hep-ph/0002072.

\bibitem{Kaidalov03}
A.~B. Kaidalov, V.~A. Khoze, A.~D. Martin, and M.~G. Ryskin,
\newblock Eur. Phys. J. {\bf C31}, 387 (2003), hep-ph/0307064.

\bibitem{HarlandLang:2010ep}
L.~A. Harland-Lang, V.~A. Khoze, M.~G. Ryskin, and W.~J. Stirling,
\newblock Eur.Phys.J. {\bf C69}, 179 (2010), 1005.0695.

\bibitem{Khoze:2001xm}
V.~A. Khoze, A.~D. Martin, and M.~G. Ryskin,
\newblock Eur.Phys.J. {\bf C23}, 311 (2002), hep-ph/0111078.

\bibitem{Harland-Lang:2013xba}
L.~A. Harland-Lang,
\newblock Phys.Rev. {\bf D88}, 034029 (2013), 1306.6661.

\bibitem{Belitsky:2005qn}
A.~V. Belitsky and A.~V. Radyushkin,
\newblock Phys.Rept. {\bf 418}, 1 (2005), hep-ph/0504030.

\bibitem{Shuvaev:1999ce}
A.~Shuvaev, K.~J. Golec-Biernat, A.~D. Martin, and M.~G. Ryskin,
\newblock Phys.Rev. {\bf D60}, 014015 (1999), hep-ph/9902410.

\bibitem{Bjorken:1992er}
J.~Bjorken,
\newblock Phys.Rev. {\bf D47}, 101 (1993).

\bibitem{Khoze:2002nf}
V.~A. Khoze, A.~D. Martin, and M.~G. Ryskin,
\newblock Eur.Phys.J. {\bf C24}, 581 (2002), hep-ph/0203122.

\bibitem{Martin:2009ku}
A.~D. Martin, M.~G. Ryskin, and V.~A. Khoze,
\newblock Acta Phys.Polon. {\bf B40}, 1841 (2009), 0903.2980.

\bibitem{Gotsman:2014pwa}
E.~Gotsman, E.~Levin, and U.~Maor,
\newblock (2014), 1403.4531.

\bibitem{Khoze:2013dha}
V.~A. Khoze, A.~D. Martin, and M.~G. Ryskin,
\newblock Eur.Phys.J. {\bf C73}, 2503 (2013), 1306.2149.

\bibitem{Khoze:2013jsa}
V.~Khoze, A.~Martin, and M.~Ryskin,
\newblock Eur.Phys.J. {\bf C74}, 2756 (2014), 1312.3851.

\bibitem{Ryskin:2009tk}
M.~G. Ryskin, A.~D. Martin, and V.~A. Khoze,
\newblock Eur.Phys.J. {\bf C60}, 265 (2009), 0812.2413.

\bibitem{Ryskin:2011qe}
M.~G. Ryskin, A.~D. Martin, and V.~A. Khoze,
\newblock Eur.Phys.J. {\bf C71}, 1617 (2011), 1102.2844.

\bibitem{Kaidalov:2003fw}
A.~Kaidalov, V.~Khoze, A.~Martin, and M.~Ryskin,
\newblock Eur.Phys.J. {\bf C31}, 387 (2003), hep-ph/0307064.

\bibitem{Khoze:2000mw}
V.~A. Khoze, A.~D. Martin, and M.~G. Ryskin,
\newblock p. 592 (2000), hep-ph/0006005.

\bibitem{Khoze:2000jm}
V.~A. Khoze, A.~D. Martin, and M.~G. Ryskin,
\newblock Eur.Phys.J. {\bf C19}, 477 (2001), hep-ph/0011393.

\bibitem{HarlandLang:2011qd}
L.~A. Harland-Lang, V.~A. Khoze, M.~G. Ryskin, and W.~J. Stirling,
\newblock Eur.Phys.J. {\bf C71}, 1714 (2011), 1105.1626.

\bibitem{Brodsky:1981kj}
S.~J. Brodsky and G.~P. Lepage,
\newblock Phys.Rev. {\bf D24}, 2848 (1981).

\bibitem{Chernyak:1981zz}
V.~L. Chernyak and A.~R. Zhitnitsky,
\newblock Nucl.Phys. {\bf B201}, 492 (1982).

\bibitem{Harland-Lang:2013bya}
L.~A. Harland-Lang, V.~A. Khoze, and M.~G. Ryskin,
\newblock (2013), 1310.2759.

\bibitem{Mangano:1990by}
M.~L. Mangano and S.~J. Parke,
\newblock Phys.Rept. {\bf 200}, 301 (1991), hep-th/0509223.

\bibitem{Parke:1986gb}
S.~J. Parke and T.~R. Taylor,
\newblock Phys.Rev.Lett. {\bf 56}, 2459 (1986).

\bibitem{Berends:1987me}
F.~A. Berends and W.~T. Giele,
\newblock Nucl.Phys. {\bf B306}, 759 (1988).

\bibitem{Heyssler:1997ng}
M.~Heyssler and W.~J. Stirling,
\newblock Eur.Phys.J. {\bf C5}, 475 (1998), hep-ph/9712314.

\bibitem{Brown:1982xx}
R.~W. Brown, K.~L. Kowalski, and S.~J. Brodsky,
\newblock Phys.Rev. {\bf D28}, 624 (1983).

\bibitem{Belanger:1992qi}
G.~Belanger and F.~Boudjema,
\newblock Phys.Lett. {\bf B288}, 210 (1992).

\bibitem{Khoze:2006um}
V.~A. Khoze, M.~G. Ryskin, and W.~J. Stirling,
\newblock Eur.Phys.J. {\bf C48}, 477 (2006), hep-ph/0607134.

\bibitem{Borden:1994fv}
D.~Borden, V.~A. Khoze, W.~J. Stirling, and J.~Ohnemus,
\newblock Phys.Rev. {\bf D50}, 4499 (1994), hep-ph/9405401.

\bibitem{Bartalini:2011jp}
P.~Bartalini {\em et~al.},
\newblock (2011), 1111.0469.

\bibitem{Jones:2013pga}
S.~Jones, A.~Martin, M.~Ryskin, and T.~Teubner,
\newblock JHEP {\bf 1311}, 085 (2013), 1307.7099.

\bibitem{Aaij:2014iea}
LHCb collaboration, R.~Aaij {\em et~al.},
\newblock J.Phys. {\bf G41}, 055002 (2014), 1401.3288.

\bibitem{HarlandLang:2012qz}
L.~A. Harland-Lang, V.~A. Khoze, M.~G. Ryskin, and W.~J. Stirling,
\newblock (2012), 1204.4803.

\bibitem{Brodsky:1981rp}
S.~J. Brodsky and G.~P. Lepage,
\newblock Phys.Rev. {\bf D24}, 1808 (1981).

\bibitem{Benayoun:1989ng}
M.~Benayoun and V.~L. Chernyak,
\newblock Nucl.Phys. {\bf B329}, 285 (1990).

\bibitem{Martin:1997wy}
A.~D. Martin and M.~G. Ryskin,
\newblock Phys.Rev. {\bf D57}, 6692 (1998), hep-ph/9711371.

\bibitem{Ronanpriv}
Ronan McNulty, private communication.

\bibitem{Beringer:1900zz}
Particle Data Group, J.~Beringer {\em et~al.},
\newblock Phys.Rev. {\bf D86}, 010001 (2012).

\bibitem{Bodwin:2007fz}
G.~T. Bodwin, H.~S. Chung, D.~Kang, J.~Lee, and C.~Yu,
\newblock Phys.Rev. {\bf D77}, 094017 (2008), 0710.0994.

\bibitem{Hoodbhoy:1996zg}
P.~Hoodbhoy,
\newblock Phys.Rev. {\bf D56}, 388 (1997), hep-ph/9611207.

\bibitem{Aaltonen:2011hi}
CDF, T.~Aaltonen {\em et~al.},
\newblock Phys.Rev.Lett. {\bf 108}, 081801 (2012), 1112.0858.

\bibitem{Martin:2009iq}
A.~D. Martin, W.~J. Stirling, R.~S. Thorne, and G.~Watt,
\newblock Eur.Phys.J. {\bf C63}, 189 (2009), 0901.0002.

\bibitem{Pumplin:2002vw}
J.~Pumplin {\em et~al.},
\newblock JHEP {\bf 0207}, 012 (2002), hep-ph/0201195.

\bibitem{Gluck:2007ck}
M.~Gluck, P.~Jimenez-Delgado, and E.~Reya,
\newblock Eur.Phys.J. {\bf C53}, 355 (2008), 0709.0614.

\bibitem{HLfut}
Harland-Lang, L. A. and Khoze, M. G. and Ryskin, M. G., future publication.

\bibitem{Harland-Lang:2013ncy}
L.~Harland-Lang, V.~Khoze, M.~Ryskin, and W.~Stirling,
\newblock Eur.Phys.J. {\bf C73}, 2429 (2013), 1302.2004.

\bibitem{Lepage:1980fj}
G.~P. Lepage and S.~J. Brodsky,
\newblock Phys.Rev. {\bf D22}, 2157 (1980).

\bibitem{Baier:1981pm}
V.~N. Baier and A.~G. Grozin,
\newblock Nucl.Phys. {\bf B192}, 476 (1981).

\end{thebibliography}
\bibliographystyle{h-physrev}

\end{document}